\documentstyle[12pt]{article}

\def\hybrid{\topmargin -20pt	\oddsidemargin 0pt
	\headheight 0pt	\headsep 0pt
	\textwidth 6.25in	
	\textheight 9.5in	
	\marginparwidth .875in
	\parskip 5pt plus 1pt	\jot = 1.5ex}

\def\baselinestretch{1.2}

\catcode`\@=11

\def\marginnote#1{}
\newcount\hour
\newcount\minute
\newtoks\amorpm
\hour=\time\divide\hour by60
\minute=\time{\multiply\hour by60 \global\advance\minute by-\hour}
\edef\standardtime{{\ifnum\hour<12 \global\amorpm={am}%
	\else\global\amorpm={pm}\advance\hour by-12 \fi
	\ifnum\hour=0 \hour=12 \fi
	\number\hour:\ifnum\minute<10 0\fi\number\minute\the\amorpm}}
\edef\militarytime{\number\hour:\ifnum\minute<10 0\fi\number\minute}
\def\draftlabel#1{{\@bsphack\if@filesw {\let\thepage\relax
   \xdef\@gtempa{\write\@auxout{\string
      \newlabel{#1}{{\@currentlabel}{\thepage}}}}}\@gtempa
   \if@nobreak \ifvmode\nobreak\fi\fi\fi\@esphack}
	\gdef\@eqnlabel{#1}}
\def\@eqnlabel{}
\def\@vacuum{}
\def\draftmarginnote#1{\marginpar{\raggedright\scriptsize\tt#1}}

\def\draft{\oddsidemargin -.2truein
	\def\@oddfoot{\sl preliminary draft \hfil
	\rm\thepage\hfil\sl\today\quad\militarytime}
	\let\@evenfoot\@oddfoot	\overfullrule 3pt
	\let\label=\draftlabel
	\let\marginnote=\draftmarginnote
   \def\@eqnnum{(\theequation)\rlap{\kern\marginparsep\tt\@eqnlabel}%
\global\let\@eqnlabel\@vacuum}  }

\def\preprint{\twocolumn\sloppy\flushbottom\parindent 2em
	\leftmargini 2em\leftmarginv .5em\leftmarginvi .5em
	\oddsidemargin -.5in	\evensidemargin -.5in
	\columnsep .4in	\footheight 0pt
	\textwidth 10.in	\topmargin  -.4in
	\headheight 12pt \topskip .4in
	\textheight 6.9in \footskip 0pt
	\def\@oddhead{\thepage\hfil\addtocounter{page}{1}\thepage}
	\let\@evenhead\@oddhead	\def\@oddfoot{}	\def\@evenfoot{} }

\def\numberbysection{\@addtoreset{equation}{section}
	\def\theequation{\thesection.\arabic{equation}}}

\def\underline#1{\relax\ifmmode\@@underline#1\else
	$\@@underline{\hbox{#1}}$\relax\fi}

\def\titlepage{\@restonecolfalse\if@twocolumn\@restonecoltrue\onecolumn
     \else \newpage \fi \thispagestyle{empty}\c@page\z@
	\def\thefootnote{\fnsymbol{footnote}} }

\def\endtitlepage{\if@restonecol\twocolumn \else \newpage \fi
	\def\thefootnote{\arabic{footnote}}
	\setcounter{footnote}{0}}  

\catcode`@=12
\relax

\def\figcap{\section*{Figure Captions\markboth
	{FIGURECAPTIONS}{FIGURECAPTIONS}}\list
	{Figure \arabic{enumi}:\hfill}{\settowidth\labelwidth{Figure
999:}
	\leftmargin\labelwidth
	\advance\leftmargin\labelsep\usecounter{enumi}}}
 \relax
\def\tablecap{\section*{Table Captions\markboth
	{TABLECAPTIONS}{TABLECAPTIONS}}\list
	{Table \arabic{enumi}:\hfill}{\settowidth\labelwidth{Table
999:}
	\leftmargin\labelwidth
	\advance\leftmargin\labelsep\usecounter{enumi}}}
 \relax
\def\reflist{\section*{References\markboth
	{REFLIST}{REFLIST}}\list
	{[\arabic{enumi}]\hfill}{\settowidth\labelwidth{[999]}
	\leftmargin\labelwidth
	\advance\leftmargin\labelsep\usecounter{enumi}}}
 \relax
\makeatletter
\newcounter{pubctr}
\def\publist{\@ifnextchar[{\@publist}{\@@publist}}
\def\@publist[#1]{\list
	{[\arabic{pubctr}]\hfill}{\settowidth\labelwidth{[999]}
	\leftmargin\labelwidth
	\advance\leftmargin\labelsep
	\@nmbrlisttrue\def\@listctr{pubctr}
	\setcounter{pubctr}{#1}\addtocounter{pubctr}{-1}}}
\def\@@publist{\list
	{[\arabic{pubctr}]\hfill}{\settowidth\labelwidth{[999]}
	\leftmargin\labelwidth
	\advance\leftmargin\labelsep
	\@nmbrlisttrue\def\@listctr{pubctr}}}
 \relax
\makeatother
\newskip\humongous \humongous=0pt plus 1000pt minus 1000pt

\newif\ifdtup

\relax
\hybrid
\def\s{\sigma}
\def\thefootnote{\fnsymbol{footnote}}
\def\be{\begin{equation}}
\def\ee{\end{equation}}
\def\ba{\begin{eqnarray}}
\def\ea{\end{eqnarray}}
\def\d{\partial}
\def\db{\bar{\partial}}

\def\S{\Sigma}

\def\a{\alpha}
\def\th{\theta}

\def\tht{\tilde{\theta}}

\def\p{\phi}
\def\e{\epsilon}
\def\square{\hbox{{$\sqcup$}\llap{$\sqcap$}}}
\def\rr{{\rm regular}}

\def\ec{\hat E^{c}_{2}}
\def\pp{\bar\p}
\def\j{\tilde J}
\def\xb{{\bar x}}
\def\px{{\vec p}}

\begin{document}
\renewcommand{\theequation}{\thesection.\arabic{equation}}
\newcommand{\beq}{\begin{equation}}
\newcommand{\eeq}[1]{\label{#1}\end{equation}}
\newcommand{\ber}{\begin{eqnarray}}
\newcommand{\eer}[1]{\label{#1}\end{eqnarray}}
\begin{titlepage}
\begin{center}

\hfill CERN-TH.7328/94\\
\hfill LPTENS-94/18\\
\hfill hep-th/9407005\\

\vskip .5in

{\large \bf Dynamical Topology Change, Compactification and Waves in
a Stringy
Early Universe\footnote{Talk presented by E. Kiritsis in the 2\`eme
Journ\'ee Cosmologie, Observatoire de Paris, 2-4 June 1994}}
\vskip .8in

{\bf Elias Kiritsis and Costas Kounnas\footnote{On leave from Ecole
Normale Sup\'erieure, 24 rue Lhomond, F-75231, Paris, Cedex 05,
FRANCE.}}\\
\vskip
 .4in

{\em Theory Division, CERN, CH-1211\\
Geneva 23, SWITZERLAND} \footnote{e-mail addresses:
KIRITSIS,KOUNNAS@NXTH04.CERN.CH}\\

\vskip .5in

\end{center}

\vskip .2in

\begin{center} {\bf ABSTRACT } \end{center}
\begin{quotation}\noindent
Exact string solutions are presented, where moduli fields are varying
with time.
They provide examples  where a
dynamical change of the topology of space is occurring.
Some other solutions give cosmological examples where some dimensions
are compactified dynamically or simulate pre-big bang type scenarios.
Some lessons are drawn concerning the region of validity of effective
theories
and how they can be glued together, using stringy information in the
region where the geometry and topology are not well defined from the
low energy point of view.
Other time dependent solutions are presented where a hierarchy of
scales is absent.
Such solutions have dynamics which is qualitatively different and
resemble
plane gravitational waves.

\end{quotation}
\vskip 2.0cm
CERN-TH.7328/94 \\
July 1994\\
\end{titlepage}
\vfill
\eject
\def\baselinestretch{1.2}
\baselineskip 16 pt
\noindent
\section{Introduction, Results and Conclusions}
\setcounter{equation}{0}

One of the strongest motivations which made (super) string theory
popular is the fact that it provides (at least perturbatively) a
consistent and maybe even UV finite theory of quantum gravity.
Moreover it unifies gravity with other
interactions including  interactions of the gauge and Yukawa type.
It is thus appropriate to try to investigate the behaviour of string
dynamics associated with gravitational phenomena. There are several
problems in gravity where the classical, and even worse the
semiclassical treatment have perplexed physicists for decades. We are
referring here to questions concerning the behaviour in regions of
strong (or infinite ) curvature with both astrophysical
(black holes) and cosmological (big-bang, wormholes) interest.
It is only appropriate to try to elucidate such questions in the
context of stringy gravity.
There has been progress towards this direction, and by now we have at
least
some ideas on how different string gravity can be from general
relativity
in regions of spacetime with strong curvatures.
We need however exact (in $\alpha'$) classical solutions of string
theory
in order to have more quantitative control on phenomena that are
characteristic of stringy gravity.

In this lecture we shall present some exact solutions to string
theory
in
3+1 dimensions with interesting intepretations like $dynamical$
$topology$ $change$, $dynamical$ $compactification$ or $dynamical$
$generation$ of $gauge$ $symmetries$ \cite{dtc} as well as wavelike
solutions \cite{w11,kk1,kk2,wave}.
The number of exact solutions to string theory where the background
fields indicate some interesting behaviour is not large, but we can
however try to get to some conclusions from studying the exact
solutions we posses.
This is the first time we have a model where we can study for example
dynamical topology
change in a region where the curvature is strong (of the order of the
Planck scale)
and where the $\alpha'$ expansion (or organization) of the effective
field theory breaks down. However we can extend at the string level
our description
past the strong curvature region (where the topology change occurs)
towards another asymptotic region where we have a different low
energy field theory.
The topology change is described by a modulus field that varies with
time.

Studies on string theory have so far given hints for the presence of
several interesting phenomena, like the existence of a minimal
distance \cite{md}, finiteness at short distances \cite{m}, smooth
topology change \cite{tc,gk}, spacetime duality
symmetries \cite{r},\cite{bu}-\cite{AG},\cite{gk}, variable
dimensionality of spacetime \cite{di}, existence of maximal
(Hagedorn) temperature and subsequent phase transitions \cite{PT}
etc.
The lessons we learn from the exact solutions we are going to
describe essentially corroborate some items on the list above and we
would like to present them in a somewhat general
context.\footnote{Ideas of a similar form have already been presented
in \cite{kk}.}

We will start from the simple case of a compactification on a flat
torus to indicate the idea and we will eventually translate for
non-flat backgrounds.
The spectrum of string physical states in a given compactification
can be generically separated in three kinds of sub-spectra.

 {\bf i})The first one consists of Kaluza-Klein-like effective field
theory modes (or momentum modes).
The masses of such modes are always proportional to the typical
compactification scale $M_{c}=1/\sqrt{\alpha'}R$ where $R$ is a
typical radius\footnote{When there are more than one compact
dimensions one can still
define the concept of a scale \cite{si}.}.

{\bf ii}) Another set contains  the  winding modes which exist here
because of two reasons. The first is that the string is an extended
object.
The second is that the target space has a non-trivial $\pi_{1}$ so
the string can wind around in a topologically non-contractible way.
In a large volume compact space, these modes are always super-heavy,
since their mass is inversely proportional to the compactification
volume $M_c$, (more precisely, it is proportional to $1/\alpha'
M_{c}$).

{\bf iii})The third class of states are purely stringy states
constructed from the string oscillator operators. Their masses are
proportional to the string scale $M_{str}=1/\sqrt{\alpha '}$.

This separation of the spectrum is strictly correct in torroidal
backgrounds.
However, further analysis indicates that one can extend
the notions of Kaluza-Klein (KK) type modes and winding modes to at
least non-flat backgrounds with some Killing symmetries.
This generalization comes with the help of the duality symmetries
present in such background fields \cite{bu}-\cite{AG},\cite{gk}.
However, the ``winding" states are not associated with
non-contractible circles of the manifold. They appear as winding
configurations in a (usually) contractible
circle associated with a Killing coordinate \cite{k2}. An example
would be given by a winding in the Cartan torus of a group, which is
however contractible inside the full group manifold.
Because of the above, such modes are special combinations of zero
modes as well
as oscillator excitations.
There may be dual versions of the model where such states are in fact
associated with non-contractible circles of the dual manifold. We
will see such a case in section 2.
Although the above seems to apply to backgrounds with Killing
symmetries, we
feel that it might be more general. There are indications
\cite{k1,k2,ko} that one can have duality symmetries without the
presence of isometries.

We will illustrate better the issues above in two cases.
The first corresponds to Cartan deformations in a WZW model.
There, although the Cartan torus is contractible, the role of
windings and momenta are played by the eigenvalues of the left and
right Cartan generators.
The duality map here involves also the oscillators (that is, the
currents).
We will analyse an example of this sort in the rest of the paper.
The second kind of behaviour concerns models of the
$SL(2,R)_{k}/U(1)$ type.
There, one has zero modes and oscillator excitations (here we mean
the non-compact parafermions), however, unlike the standard case, the
energies associated to such oscillators are $k$-dependent.
Duality interchanges zero modes with energies $~1/k$ with high lying
modes
with energies $\sim k$ \cite{k2}.

Once we have the picture above, concerning different types of string
excitations we can state that the notion of a string effective field
theory makes sense  when the ``winding" states as well as the
oscilator states are much heavier than the field theory-like KK
states.
Here, we will assume the presence of a single scale (except
$\alpha'$).
If there are more such scales then one has to investigate the
different regimes. In any such regime, our discussion below is
applicable.

Using the $\alpha'$-expansion one finds the effective theory which is
applicable to low energy processes ($E_t<M_{st}$) among the
massless states, as well as the lowest lying  KK states, provided,
that the typical mass scale $M_c$, (which could be a compactification
scale, gravitational curvature scale etc ) is much below the string
scale $M_{st}$.
In the last few years, a lot of activity was devoted in understanding
the effective theories of strings at genus zero, \cite{g0} and in
some cases the genus one corrections were included \cite{g1}. The
output of this study confirms that the winding and the ${\cal
O}(M_{st})$ string  superheavy modes can be integrated out and one
can  define consistently  the string  low energy theory in terms of
the massless and lowest massive KK states. Thus, the perturbative
string solutions are well  described by classical gravity coupled to
some gauge and matter fields with unified gauge, Yukawa and self
interactions. As long as  we stay in the regime where $E_{t}, \tilde
M_c <M_{st}$, our
description of the physics in terms the the effective theory is good
with well defined and calculable ${\cal O}(\frac{E_t}{M_{st}}),{\cal
O}(\frac{M_c}{M_{st}})$ corrections.

The situation above  changes drastically once the mass of  "winding"
type  states becomes smaller than the string scale and when (usually
at the same time) the KK modes have masses above the string scale.
This is the case when the typical $M_c$ scale is larger than
$M_{st}$. When this occurs, the relevant modes are not any more the
KK ones but rather the winding ones. Thanks to the well known by now
generalized string duality \cite{bu}-\cite{AG},\cite{gk} (e.g. the
generalization of the well known R to 1/R torroidal duality) it is
possible to find an alternative effective field theory description
by means  of ${\cal O}(\frac{E_{t},\tilde M_c }{M_{st}})$ dual
expansion, where one uses the dual background which is characterized
by $\tilde M_c$ instead of the initial one characterized by a high
mass scale. The dual mass scale  $\tilde M_c=\frac{M^{2}_{st}}{M_c}$
is small when the $M_c$ becomes big and vice-versa. We observe that
in both extreme cases, either
(i) $M_c < M_{st}< \tilde M_c$ or
(ii) $M_c>M_{st}>\tilde M_c$,  a field theory description exists in
terms of the original curved background metric in the first case or
in terms of its dual in the second case. This observation is of main
importance since it extends the notion of the effective field
theories in backgrounds with associated high mass scales (due to size
or curvature).

Then,
a general string solution would give rise to a set of effective field
theories defined in restricted regions of space-time $(x^{\mu})_I$,
I=1,2,3... with $M_I$  smaller or of the same order as $M_{st}$; If
$T_{I,J}$ is the boundary region among $(x^{\mu})_I$ and
$(x^{\mu})_J$, then on $T_{I,J}$ we have almost degenerate effective
characteristic scales $M_I\sim M_{st}\sim M_J$.
In such regions, the effective field theory description of regions
$I,J$ break
down (individually), and the full string theory is needed in order to
have a smooth transition between the two.
A goal in that direction would be to establish some simple rules that
would
provide the extra (stringy)\footnote{Here by stringy we mean exact to
all orders in $\alpha'$, or equivalently the full CFT description.}
information that would patch the field theoretic regions together.
This effectively amounts to a reorganization of the $\alpha'$
expansion.\footnote{In simple backgrounds, like torroidal ones this
patching can be effectively done \cite{gp} by constructing an
effective action, containing an infinite number of fields.}
The models we are proposing in this paper could very well
serve as a laboratory towards answering this question.
This is certainly important, since many interesting phenomena happen
precisely
at such regions in string theory. We can mention, the gluing of
dual solutions in cosmological contexts, \cite{bv} and global
effective theories for large regions of internal moduli spaces.

A related issue here is, that with each region, one has an associated
geometry
and spatial topology, as dictated by the effective field theory.
It turns out that moving from one region to another not only the
geometry can change but also the topology. Examples in the context of
Calabi-Yau compactifications
\cite{tc} and more simple models, \cite{gk} have been given.
There is another important point about topology change that we would
like
to stress here. Where topology change happens, depends crucially on
the values of some of the parameters in the background. One such
parameter is always $\alpha'$,
but usually, in string backgrounds, there are others, like various
different levels for non-simple WZW and their descendant CFTs,
various radii or related moduli etc.
The absolute judge concerning topology change is the effective field
theory.

As we will see in later sections, the solutions we will describe can
be viewed as a time dependent background, where a modulus of space
(or internal space) changes with time\footnote{Some solutions of that
type were recently described in \cite{tse}.}.
For the solution that describes topology change we have
at $t=0$ a manifold with the topology of a disk times a line
which evolves at $t=\infty$ to a manifold with the topology of a
cylinder
times a line.
The topology changes in the intermediate (stringy) region.
Of course, the backgrounds we are describing are time-dependent and
thus,
strictly speaking, we loose the meaning of energy (or mass in the
compactified case). However, we will show that there are regimes
where the time dependence
is adiabatic, and where it makes sense to use an approximate concept
of energy
or mass. Solutions with time dependent radii have been found in the
past, solving the string equations to leading order in $\alpha'$
\cite{r1}. The advantage of the solutions we present is that they are
exact string solutions (to all orders in $\alpha'$).

In this context, we will be able to demonstrate our general picture
described above. One of the models considered, is $SL(2,R)_{k}$ or
$SU(2)_{k}$ with
its Cartan deformed. This deformation is parametrized by a new
continuous parameter, $R$ such that at $R=1$ we have the $SU(2)$ or
$SL(2,R)$ model.
The SL(2,R) family of models was considered in \cite{gk} as a simple
example
of topology change.
The topology change happens only at the boundary
$R=0,\infty$.\footnote{In \cite{gk} a variation was given where the
topology change happens in the interior of the $\s$-model moduli
space.}
Before we consider any time variation of $R$, let us illustrate a few
features of the (compact) static model.
The spectrum has the form \cite{sky}:
\be
L_{0}+\bar L_{0}=2{j(j+1)\over k+2}-{m^2+\bar m^2\over k}+{1\over
2k}\left[(m-\bar m+kM)^2
R^2+{(m+\bar m+kN)^2\over R^2}\right]+N'+\bar N'\label{1x}
\ee
where $m,\bar m\in [0,2k]$, $M,N,N',\bar N'\in Z$.
For $R\sim 1$ the low-lying spectrum is close to that of $S^{3}$.
The effective field theory contains only the low-lying spectrum  and
the geometry is certainly that of $S^{3}$.
However, for $R>>1$, the low lying spectrum is dominated from the
Cartan contributions and the geometry, from the effective field
theory point of view
is that of $S^{1}$. The leftover piece, which is that of
$SU(2)_{k}/U(1)$
is strongly curved (for $k\sim {\cal O}(1)$) and thus not visible at
low energy.
Thus, we see that the
change
of effective field theory happens in the $R>>1$ or $R<<1$ regions and
is certainly
not describable in the language field theory.
In the half line $0<R<\infty$, we need one effective
field theory to describe the region for $R<<1$ and $R>>1$ and another
in-between.

The modulus $R$ can be made to be time dependent without destroying
the
exactness of the solution. Moreover as we show in the main text there
are regions where this dependence is adiabatic.
Thus, in the (euclidean) $SL(2,R)_{k}$ case, we can make the topology
change picture described above, dynamical.

We also find solutions in which spacetime starts out (at $t=0$) as 5
(or more) dimensional but the extra dimension becomes compact and
settles at $t=\infty$ to the $self-dual$ point $R=1$, of Plankian
size.
In such a case the early universe has at late times an extra SU(2)
local gauge symmetry generated by the dynamical compactification.

Another application of such solutions could be in considering strings
at finite temperature.
They would describe a string ensemble with temperature that varies
(adiabatically) in space. It might be interesting to entertain such
an idea in more detail, in order to investigate temperature gradients
in string theory.
We will comment on this possibility in the last section.

There are on the other hand exact string solutions, with non-trivial
background
fields, which behave as flat space, in the sense that there is only a
unique scale, and this is $\alpha'$. Such exact backgrounds have been
analysed recently \cite{w11,kk1,kk2,wave} and resemble plane
gravitational waves.
We will analyse the simplest of these models in order to show the
qualitative differences from the models described above and point out
some interesting dynamics that might arise in such stringy
backgrounds.

\section{The time-independent model}
\setcounter{equation}{0}

We would like to construct exact classical solutions to string theory
where
some moduli vary with time and their variation can induce geometry or
topology
change, in the sense presented in the introduction.
In order to do that, we will start from a solution where the modulus
is time independent, but arbitrary (alternatively speaking, there is
no potential for it, in the effective action).
This will be the subject of this section where we will use the
results of
\cite{gk}. In the next section we will proceed to include time
dependent moduli.

The models we will analyse are related with the Cartan deformations
of SU(2)$_{k}$ or SL(2,R)$_{k}$ and its euclidean continuation
$H^{3}_{+}$. We will briefly describe the
$\s$-model action for this radius deformation.
For more details we refer the reader to \cite{gk}.

Although the deformation of SU(2)$_{k}$ was originally found using
O(2,2,R)
deformations, \cite{hs}\footnote{The fact that O(2,2,R)
transformations can produce marginal current-current perturbations
infinitesimally was observed in \cite{k2}.}
we will present a different approach \cite{gk}, which has the
advantage of showing that the $\s$-model action and dilaton we will
thus obtain, are exact to all orders in the $\alpha'$ expansion (in
some scheme).

Conformal perturbation theory indicates that there is a line of
theories
obtained by perturbing around the $SU(2)$ or $SL(2,R)$ WZW model  by
$\int J^{3}{\bar J}^{3}$.
The theories along the line have a $U(1)_{L}\times U(1)_{R}$ chiral
symmetry.
Thus the $\sigma$-model action of these theories must satisfy at
least the following four properties:
\vskip .6cm

{\bf 1}) It should have $U(1)_{L}\times U(1)_{R}$ chiral symmetry
along the
line.

{\bf 2}) It should have the group property: $\delta S \sim \int
J^{3}{\bar
J}^{3}$
at {\em any} point of the line.

{\bf 3}) At a specific point it should reduce to the known action of
the
   $SU(2)$ or $SL(2,R)$ WZW model.

{\bf 4}) The $\s$-model  should be conformally invariant.

\vskip .6cm

In \cite{gk} it was shown that properties (1-3) above imply that the
$\s$-model action is:
\be
S(R)={k\over 2\pi}\int d^2
z\left({(1-\S)\d\th\db\th+R^2(1+\S)\d\tht\db\tht
+(1+\S)(\d\th\db\tht-\d\tht\db\th)\over 1+\S+R^2(1-\S)}+\d x\db
x\right)\label{su1}
\ee
where the angles $\th,\tht$ take values in $[0,2\pi]$.
For SU(2), $\S=\cos 2x$ with $x\in [0,\pi/2]\cup [\pi,3\pi/2]$
whereas for
$H_{3}^{+}$, $\S=\cosh 2x$ with $x$ real.
Finally we can go to SL(2,R) by changing the sign of the $\d x\db x$
term in
(\ref{su1}).
R is the parameter which varies along the line, $R\in [0,\infty)$.

So far we have not imposed conformal invariance.
At one-loop the $\beta$-function equations determine the dilaton,
which has not been fixed yet:
\be
\p(x,R)=\log[1+{1-R^2\over 1+R^2}\S(x)]+f(R).\label{dil1}
\ee
where $f(R)$ is an arbitrary R-dependent constant.
The only way (\ref{su1}) can change consistent with our requirements
(1-3)
is by a redefinition of $R$, which implies that there is a scheme in
$\sigma$-model perturbation theory where the metric and the
antisymmetric tensor receive no higher order corrections in $\a'$.
In such a case also the dilaton receives no higher order
corrections.\footnote{The dilaton $\beta$-function (central charge)
does get corrections.
This is what is happening also in the WZW model.
However, like in that case, one can replace $k$ with $k+2$ in front
of
the action.
Then the central charge is given by the classical and 1-loop piece
only, without spoiling the vanishing of the other
$\beta$-functions.}.

The $R$-dependent constant in (\ref{dil1}) can be fixed by the
requirements that
$\sqrt{G(R)}e^{\phi(R)}$ (which represents the physical string
coupling) is invariant along the line.
This implies that $f(R)=\log(R+1/R)$ and that
\be
\p(x,R)=\log[(1-\S(x))R+(1+\S(x))/R]+\phi_{0}.\label{dil}
\ee
The constant $\phi_{0}$ in (\ref{dil}) is $R$-independent.
When $R\to 0,\infty$, the dilaton has the appropriate asymptotic
behaviour.
This is an independent way of fixing the constant R-dependent part of
the dilaton and the result coincides with the previous method.

Several observations are in order here, \cite{gk}.

$\bullet$ For the line of theories
above one has $R\to 1/R$ duality.
If we parametrize $R$ around the WZW point $R=1$ as
$R=1+\epsilon+{\cal O}(\epsilon^2)$ then the duality transformation
becomes $\epsilon\to -\epsilon +{\cal O}(\epsilon^2)$. This is
equivalent to the Weyl invariance ($J^{3}\bar J^{3}\to -J^{3}\bar
J^{3}$) of the theory at $R=1$.
The fact that the remnants of the Weyl invariance can be retained far
away
is guaranteed by the fact the perturbation theory converges and it
can also organized in order to preserve duality order by order.
Non-perturbative effects (in $\alpha'$) exist only for SU(2) and
there a glimpse at the exact partition function \cite{sky} indicates
that they don't spoil the symmetry.

$\bullet$ At $R=0$ the model becomes a direct product of a
non-compact boson
and the vector coset $SU(2)_{k}/U(1)_V$
(or $SL(2,R)_{k}/U(1)_V$).
Strictly speaking, the radius of $\tht$ becomes zero, but this is
equivalent
to a non-compact boson (as can be verified from the exact partition
function in the $SU(2)$ case).
There is also a constant antisymmetric tensor piece, but in the limit
it can be safely dropped since it couples to a non-compact
coordinate.

$\bullet$ At $R=\infty$ the model becomes a direct product of a
non-compact boson ($\th$, this time) and the axial coset
$SU(2)_{k}/U(1)_A$ (or $SL(2,R)_{k}/U(1)_A$).

$\bullet$ The three observations above imply that the axial and the
vector
cosets are equivalent CFTs.

We will look now at the geometry along the line, \cite{gk}.
The $\sigma$-model metric in the case of deformed
$SU(2)$ (in the coordinates $\th$, $\tht$, $x$)
is given by
\be
G\sim k\left(
\matrix{{\sin^2 x\over  \cos^2 x +R^2 \sin^2 x}& 0&0\cr
0&{R^{2}\cos^2 x\over \cos^2 x+R^2 \sin^2 x }&0\cr
0&0&1\cr}\right).\label{metric}
\ee
The scalar curvature ${\hat R}$ is
\be
{\hat R}=-{2\over k}{2-5R^2 +2(R^{4}-1)\sin^2 x \over
 (1+(R^2-1)\sin^2 x)^2} \; .
\label{r}
\ee
The manifold is regular except at the end-points where
\be
{\hat R}(R=0)=-{4\over k\cos^2 x}\;\;,\;\;{\hat R}(R=\infty)=-{4\over
 k\sin^2 x}\; .
\ee
At $R=1$ we get the constant curvature of $S^{3}$, ${\hat R}=6/k$.

It should be noted that the geometric data (metric, curvature, etc.)
are
invariant under $R\rightarrow 1/R$ and $x\rightarrow \pi/2-x$.
To put it differently the two (dual) branches are diffeomorphic.
Another interesting object is the volume of the manifold as a
function of $R$
that can be computed to be
\be
V(R)\sim {R\;{\log}R\over R^{2}-1}\label{volume}
\ee
satisfying $V(R)=V(1/R)$. The volume becomes singular
only at the boundaries of moduli space, $R=0,\infty$.
To be more precise, it vanishes there, reflecting the vanishing of
the radius
of one of the angles (either $\th$ or $\tht$).
It is interesting to note though that at the limits, the model
factorizes
to a theory that has zero volume (corresponding to one of the angles)
and another with infinite volume (the coset, this is due to the
semiclassical singularity).
The zero volume space dominates the infinite volume one.
The ``string" volume, $\int e^{\phi}\sqrt{G}$ is constant along  the
line.

The topology of the manifold is $S^3$ except at the end-points.
Since one of the circles there shrinks to a point
we have really a degeneration  of the manifold to a disk times a
point.

For $H^{+}_{3}$, the trigonometric functions in (\ref{r}) are
replaced
by the
corresponding hyperbolic functions. Here the manifold has a curvature
singularity for $0\leq R <1$.
It should be noted that here the two dual branches are not
diffeomorphic
to each other. This is obvious, since one branch is singular and the
other is not.
The transition point is at $R=1$, where a singularity ``sneaks in"
from infinity.
In this case the manifold has always infinite volume.

The $R$ marginal deformation generates a continuous  family of CFTs
that
interpolate between two manifolds with  different topology: the
``cigar"
shape ($R=\infty$) with the topology of the disk and the ``trumpet"
shape
($R=0$) with the topology of a cylinder.
The essential transition happens at $R=1$.
In the $SL(2,R)$ case similar things happen as above.

In \cite{gk} we have given also a continuous family of theories,
where the topology change happens in the interior of moduli space.
There, for example, an $S^3$ would evolve into a disk times a finite
radius circle. We will not pursue further these backgrounds here.

\section{Dynamical Topology Change}
\setcounter{equation}{0}

We would like to make the static picture described in the previous
section
dynamical, ie. evolving in real time.
In order to do that, we would need to add time into the $\s$-model
(\ref{su1}).
Thus, we will consider the following 4-d $\sigma$-model action:
$$
S_{3+1}=S(R(t))-{1\over 2\pi}\int d^2 z\; \d t\db t-{1\over 8\pi}\int
d^2z \;R^{(2)}\Phi(x,t)=
$$
\be
={k\over 2\pi}\int d^2
z\left({(1-\S)\d\th\db\th+R^2(t)(1+\S)\d\tht\db\tht
+(1+\S)(\d\th\db\tht-\d\tht\db\th)\over 1+\S+R^2(t)(1-\S)}+\d x\db
x\right)-\label{3+1}
\ee
$$
-{1\over 2\pi}\int d^2 z\; \d t\db t-{1\over 8\pi}\int d^2z
\;R^{(2)}\Phi(x,t)
$$
where $S_{1+2}(R(t))$ is the action in (\ref{su1}) but where the
radius has  (an unspecified for the moment) time dependence.
We would like to find exact string solutions of the form (\ref{3+1}).

Let us consider the following exact CFT described by the action
$\tilde S=\tilde S_{1}+\tilde S_{2}$
with
\be
\tilde S_{1}={k'\over 2\pi}\int d^2 z\;\left[-\d t\db t +{\tan^2
t\over
k}\d\th\db\th\right]-{1\over 8\pi}\int d^2
z\;R^{(2)}\log[\cos^2 t]\label{s11}
\ee
\be
\tilde S_{2}={k\over 2\pi}\int d^2 z\;\left[\d x\db x +{1+\S(x)\over
1-\S(x)}\d\phi\db\phi\right]-{1\over 8\pi}\int d^2
z\;R^{(2)}\log[1-\S(x)]\label{s22}
\ee
with $\phi=\tht+\th/k$.
This is (almost) a product of the (minkowskian analytic
continuation\footnote{This is done by rotating the killing coordinate
on the imaginary axis and changing the sign of $k'$}) of the
$SU(2)_{k'}/U(1)$ coset model (described by action $\tilde S_{1}$)
and
of the $SU(2)_{k}/U(1)$ or the $SL(2,R)_{k}/U(1)$ coset model (this
depends on the form of $\S(x)$.
The coupling between the two models will be discussed in detail
below.

Let us do a duality transformation on the action $\tilde S$ with
respect to the angle $\th$. Then we obtain precisely the $\s$-model
action (\ref{3+1})
with
\be
R(t)=\sqrt{k}\tan(t/\sqrt{k'})
\ee and
\be
\Phi(x,t)=\log\left[(1+\S(x))\cos^2(t/\sqrt{k'}+k(1-\S(x))\sin^2(t/\sqrt
{k'})
\right]+{\rm constant}\label{dila2}
\ee
We have succeded to find an exact CFT whose (dual) $\s$-model
interpretation
is that of a CFT with a modulus varying with time.
By considering other tensor products using flat space or the
$SU(2)/U(1)$
coset we can construct more exact solutions with different types of
time dependence. We will now show that this is all there concerning
the time
dependence of $R(t)$.

To see this we will solve the (1-loop) $\beta$-function equations for
the $\s$-model action (\ref{3+1}).
The one-loop $\beta$-function equations imply
\be
{R'''\over R''}={R''\over R'}+{R'\over R}\label{R}
\ee
for $R(t)$
\be
\Phi(x,t)=\log\left[{1+\S(x)+R^{2}(t)(1-\S(x))\over
R'(t)}\right]+{\rm constant}\label{dila}
\ee
for the dilaton and
\be
\delta^{(1)}c=-{6\over k}+{3\over 2}{R''\over R'}\left(2{R'\over
R}-{R''\over R'}\right)
\ee
for the one-loop correction to the central charge.
It should be noted that (\ref{R}) is invariant under $R(t)\to
1/R(t)$.

Equation (\ref{R}) has three classes of solutions corresponding to
flat space, SU(2)/U(1)
or SL(2,R)/U(1) coset models and their duals.
In more detail, (\ref{R}) can be integrated to
\be
R'=C_{1}R^2+C_{2}\label{R2}
\ee
$1/R(t)$ is a solution of the same equation with $(C_{1},C_{2})\to
(-C_{2},-C_{1})$.
The solutions are (up to shifts in $t$):

({\bf ia})  $C_{1}=0$:
\be
R^2(t)=C_{2}^{2}t^2\label{ia}
\ee
For $t\in [0,\infty)$, $R^2\in [0,\infty)$.

({\bf ib})  $C_{2}=0$:
\be
R^2(t)={1\over C_{1}^2t^2}\label{ib}
\ee
For $t\in [0,\infty)$, $R^2\in [0,\infty)$.

({\bf ii}) $C_{1}C_{2}>0$:
\be
R(t)=\sqrt{C_{2}\over C_{1}}\tan(\sqrt{C_{1}C_{2}}t)\label{ii}
\ee
For $t\in [0,\pi/2\sqrt{C_{1}C_{2}}]$,  $R^2\in [0,\infty)$.

({\bf iii}) $C_{1}C_{2}<0$:
\be
R(t)=\sqrt{-{C_{2}\over
C_{1}}}\tanh(\sqrt{|C_{1}C_{2}|}t)\;\;\;\;\;{\rm
and}\;\;\;\;\;R(t)=\sqrt{-{C_{2}\over
C_{1}}}\coth(\sqrt{|C_{1}C_{2}|}t)\label{iii}
\ee
Here, for the $tanh$ solution, $R^2\in [0,1]$ whereas for the $coth$
solution
$R^2\in [1, \infty )$.

For all of the above $\delta^{(1)}c=\mp{6\over k}+4C_{1}C_{2}$, where
the $-$ and + corresponds to SU(2) and SL(2,R).

\setcounter{footnote}{0}
We can also discuss here the adiabaticity conditions for the
solutions above.
The backgrounds we describe change with time so, strictly speaking
there is no conserved energy. However if there is an adiabatic region
where masses change
slowly then one could still use the concept of energy to a good
approximation.
The adiabaticity condition, as we will show later on is $|R'/R|<<1$
which using
(\ref{R2}) becomes $|C_{1}R+C_{2}/R|<<1$.

For (i) the adiabatic region is at $t>>1$.

In case (ii) which corresponds to $SU(2)/U(1)$ in the dual version,
we obtain that $|R'/R|\geq 2\sqrt{C_{1}C_{2}}$.
Thus, there is always a lower bound in adiabaticity which tends to
zero
only if $k\to \infty$, where $k$ is the central element of the
$SU(2)_{k}/U(1)$
theory.

Finally in case (iii) the adiabatic region is again $t>>1$.
In cases (ii) and (iii) the ``size" of the adiabatic region is
governed by $|C_{1}C_{2}|$,
in the sense that the adiabaticity condition implies
$\sqrt{C_{1}C_{2}}t>>1$.
We can blow-up that region by making $|C_{1}C_{2}|$ arbitrarily
small.
In fact, we can even arrange for the total system to have $c=4$ and
still be able to blow up the adiabatic region.

As we will now show, the presence of the solutions (i-iii) is not
accidental.
If we do a duality transformation in (\ref{3+1}) with respect to the
$\th$ angle
we obtain $\tilde S=\tilde S_{1}+\tilde S_{2}$
with
\be
\tilde S_{1}={1\over 2\pi}\int d^2 z\;\left[-\d t\db t +{R^2(t)\over
k}\d\th\db\th\right]+{1\over 8\pi}\int d^2
z\;R^{(2)}\log[R'(t)]\label{s1}
\ee
\be
\tilde S_{2}={k\over 2\pi}\int d^2 z\;\left[\d x\db x +{1+\S(x)\over
1-\S(x)}\d\phi\db\phi\right]-{1\over 8\pi}\int d^2
z\;R^{(2)}\log[1-\S(x)]\label{s2}
\ee
with $\phi=\tht+\th/k$. Using this redefinition $\phi$ is an
independent
angle, so one might think that the two models are decoupled. This is
not the case though due to the presence of $1/k$ in the redefinition.
We will discuss this coupling in more detail below.
$\tilde S_{2}$ describes the $SU(2)/U(1)_{V}$ or $SL(2,R)/U(1)_{V}$
coset.

Of course, in this ``dual" formulation, we know the exact conformal
field theory associated to $\tilde S_{1}$
corresponding to the three types of solutions found above.
Case (i) is flat 2-d space (we could also have a linear dilaton but
we will
not entertain this further). Case (ii) corresponds to the Minkowski
version of the $SU(2)_{k'}/U(1)$ coset model, (with $k'\sim
1/C_{1}C_{2}$).
Finally case (iii) corresponds to the 2-d black-hole \cite{w1}.

Another way to state what was said  above is that there  is an O(2,2)
transformation \cite{mv,gpp} which maps the model (\ref{3+1}) to a
product of two 2-d  target spaces.\footnote{In fact the solutions we
describe
are special lines in the general O(2,2) moduli space described in
\cite{gpp}.}
Start from the action (\ref{3+1}) and perform the following O(2,2)
transformation:
$M=\left(\matrix{a&b\cr c&d\cr}\right)$ with
\be
a=\left(\matrix{0&-1\cr
1&0\cr}\right)\;\;\;,\;\;\;b=\left(\matrix{0&0\cr
0&0\cr}\right)\;\;,\;\;
c={1\over k}\left(\matrix{1&0\cr
0&1\cr}\right)\;\;\;,\;\;\;d=\left(\matrix{0&-1\cr 1&0\cr}\right)
\ee
We thus obtain (\ref{s1},\ref{s2}) with an extra duality
transformation on $\tht$ which gives instead of $\tilde S_{1}$:
\be
\tilde S'_{1}={1\over 2\pi}\int d^2 z\;\left[-\d t\db t +{k\over
R^2(t)}\d\tht\db\tht\right]-{1\over 8\pi}\int d^2
z\;R^{(2)}\log[R^2/R']
\ee
The coupling between the original models is hidden here in the fact
that the $O(2,2)$ matrix has fractional elements.

To summarize: in the dual formulation, the original theory
corresponds to the product of two conformal fields theories, one
being $SU(2)_{k}/U(1)$ or $SL(2,R)_{k}/U(1)$ (depending on what
$\S(x)$ is)  and the other (giving the time dependence) is one of the
 $SU(2)_{k}/U(1)$,
$SL(2,R)_{k}/U(1)$, flat 2-d space or its dual.

As it is obvious from (\ref{s2}), the two CFTs corresponding to
$\tilde S_{1}$
and $\tilde S_{2}$ are coupled. The type of coupling that exists is
precisely the one which transforms a parafermionic theory times a
free boson
to an $SU(2)_{k}$ theory.
This is a direct coupling of the parafermionic $Z_{k}$ charge to the
translational $Z_{k}$ charge of the compact boson.
In terms of the $\s$-model angles, a translation of $\tht$ by $2\pi$
induces a translation of $\phi$ by $2\pi/k$.
This implies that the operators in the spectrum of the theory must
have the mod-k residues of their angular momenta (corresponding to
$\tht$ and $\phi$) the same.
For scattering amplitudes,  this just affects only the way one
couples the operators of the two theories.

We can also write now the exact central charge of the model:
\be
c=2{k+1\over k-2}+2{3+2C_{1}C_{2}\over 3-4C_{1}C_{2}}\label{c}
\ee
Observe that we can choose $C_{1}C_{2}=3/2(4-k)$ so that the total
central charge is exactly 4. This is important because, if the
central charge differs substantially from four, we loose the clear
four-dimensional interpretation, whereas if it close to four then the
extra compactified theory will have many low lying states which again
will interfere with those coming from the four dimensional part. When
$c=4$, the background admits N=4 superconformal symmetry \cite{koun}.
Observe also that we can have this extra symmetry, and at the same
time, by choosing $k\to \infty$, be in the semiclassical regime,
where time evolution is adiabatic for a long time.

There are also some global possibilities to be addressed.
Let us first start from the Euclidean case ($t\to it$).
There, we have two options: $SU(2)$ or $SL(2,R)$ (and the Euclidean
version, $H^{+}_{3}$).
We will focus first  $SU(2)_{k}$ in the sequel although using also
$SL(2,R)$
gives some interesting cosmological models with a non-compact
spatial slice and dynamical topology change.
Concerning the time dependence of the radius $R(t)$, we have the
three different cases, (i,ii,iii).

Once we go to the Minkowski case, there are several possibilities.
We will consider first the case where $t$ is the time.
If we look at the dual action (\ref{s1}) then it is obvious that
there are
two distinct possibilities for the coordinate $\tht$: that it has a
finite radius, or that it is a non-compact coordinate.
The former case is a suitable orbifold of the latter with respect to
a discrete
infinite abelian group.

There is another possibility in which $x$ is taken as time, by doing
the $x\to ix$ continuation in the Euclidean case.
For the possibilities (ii,iii) this model has a similar
interpretation
as before with $t\leftrightarrow x$.

We will study here the semiclassical picture of these solutions.
If we consider the time-independent case, for the deformed
$SU(2)_{k}$
theory at radius $R$, it is not difficult to show that the
eigenfunctions of the Laplacian (specified by the metric
(\ref{metric}) and the dilaton (\ref{dil}) as $\square_{3}\equiv
{e^{-\Phi}\over G}\d_{\mu}e^{\Phi}G^{\mu\nu}\d_{\nu}$) are the same
as those for SU(2), namely the standard D-functions, $D^{j}_{m,\bar
m}
(x,\th,\tht)=e^{i(m-\bar m)\th/2+i(m+\bar m)\tht/2}d^{j}_{m,\bar
m}(x)$.
The energy ($L_{0}+\bar L_{0}$) eigenvalues change though:
\be
E_{3}(R)={1\over k}\left[2j(j+1)-m^2-\bar m^2+{1\over 2}(m-\bar m)^2
R^2+{1\over 2}{(m+\bar m)^2\over R^2}\right]+{\cal
O}(1/k^2)\label{e3}
\ee
in agreement with the exact formula, (\ref{1x}) with $M,N,N',\bar
N'=0$.
If we now consider the four-dimensional case then we obtain that the
4-d stringy Laplacian is given as
\be
\square_{4}={\d^2\over \d t^2}+(R'/R-R''/R'){\d\over \d
t}+\square_{3}\label{box}
\ee
Changing variables from $t\to R(t)$, and separating variables we
obtain that the wavefunctions can be written in the form $F_{j,m,\bar
m}(R)D^{j}_{m,\bar m}
(x,\th,\tht)$ where $F(R)$ satisfies the following second order
differential equation:
\be
F''+{1\over R}F'+{E_{4}-E_{3}(R)\over
(C_{1}R^2+C_{2})^2}F=0\label{eq}
\ee
where $E_{3}(R)$ is given in (\ref{e3}) and we used (\ref{R2}).
The behaviour of solutions varies for cases (i-iii).
In case (i) it is easy to see from  (\ref{eq}) that the spectrum has
a continuum part (when either $m\pm \bar m=0$ is valid) whose
wavefunctions are Bessel functions as well as a discrete part whose
wavefunctions are localized at $t=0$.

For cases (ii,iii) the wavefunctions are hypergeometric function in
the variable $x=-C_{1}R^2/C_{2}$.
$x\geq 0$ corresponds to case (iii) while $x\leq 0$ to case (ii).
The spectrum in both cases has continuous and discrete parts.
These wavefunctions correspond to specific oscillator states in the
dual coset model (\ref{s1}).

The semiclassical wavefunctions described above give a picture of the
zero mode geometry which however does not reflect the behaviour of
the whole spectrum.
This can be seen by looking at the zero mode spectrum in the dual
formulation
(\ref{s1},\ref{s2}).
The Laplacian here factorizes into a sum of
\be
{\d^2\over \d t^2}+(R'/R+R''/R'){\d\over \d t}+{1\over R^2}{\d^2\over
\d \th^2}
\label{box2}
\ee
and the laplacian of the $SU(2)/U(1)$ coset.
Their wavefunctions though, are coupled as was mentioned earlier.
Observe that although (\ref{box}) is invariant under $R(t)\to
1/R(t)$,
(\ref{box2}) is certainly not.

In case (i), for example, we have almost the same wavefuctions on the
spatial slice (although the $d$-functions are vector ($m=\bar m$))
but the time dependent part of the wavefunction has continuous
spectrum only
and behaves as a plane wave for large $t$. This case has been
analysed in detail in \cite{k2}.
For case (ii) we obtain similar results as in the original version
while in case (iii) again we have different zero mode spectrum.

For fixed time, the transverse spectrum of energies (or masses)
behaves as $E\sim m^2/R^2+n^2R^2$. The adiabaticity condition is
essentially $|\dot
E/E|<<1$ which translates to $|\dot
R(m^2/R^2-n^2R^2)/R(m^2/R^2+n^2R^2)|<<1$
and eventually to $|\dot R/R|<<1$ as advocated earlier.

\section{The Physical Interpretation}
\setcounter{equation}{0}

In this section we will discuss in more detail the physical
interpretation
of some of the models presented in the previous section.

{\bf A}) Case (iii) time dependence associated with deformed $SU(2)$
($\S(x)=\cos 2x$).

We have chosen $C_{1}=-C_{2}=\sqrt{3/2k'}$ in order to have the
standard
SL(2,R) model.\footnote{Since $C_{2}/C_{1}$ determines the radius of
$\tht$
we can go to other rational values by orbifolding.}
Here in terms of the radial time dependence we have two branches (see
 (\ref{iii})).
However both of them participate in this Minkowski signature
solution.
The best way to see that is to write the dual model in light-cone
like coordinates. The action (\ref{s1}) is precisely that of the 2-d
black hole
\cite{w1}, with metric $ds^2=k'dudv/(1-uv)$ and the two branches in
(iii) are realized in regions I and V while there is case (ii), that
is the Minkowski
version of SU(2)/U(1) that appears in between, \cite{g}.
Thus the evolution of the radius has three patches.
In the first (corresponding to region I), $R(t)\sim tanh\;t$.
At $t\to \infty$, $R(t)\equiv 1$. The effective string coupling is
given by
\be
g^{2}_{eff}=e^{-\phi}={e^{-\phi_{0}}R'(t)\over
1+\S(x)+R^{2}(t)(1-\S(x))}\label{gstr}
\ee
where we used (\ref{dila}) and where $\phi_{0}$ is the arbitrary
constant part of the dilaton.
Thus, at $t\to\infty$, $g_{eff}\to 0$ exponentially fast.
At $t\to 0$, $R(t)\sim t$ and the string coupling is finite and given
by
the (anisotropic) $SU(2)/U(1)$ dilaton.
At this point the time dependence is matched to case (ii) with $t=0$
and continues till $t\to \pi\sqrt{k'/6}$ where $R(t)\sim \infty$ and
$g_{eff}$ is finite. The adiabatic period in this model is around the
"horizon" in dual model.
Finally, from that point on, the $t$-dependence is matched into the
$coth$ branch at $t=0$ with $R(t)\sim 1/t$ and evolves till $t\to
\infty$, where $R(t)\sim 1$ and the coupling is again exponentially
small.

As argued in the previous section, we will adjust $k+k'=4$ so that
the central charge is $c=4$.
If $k>>1$ then $S^3$ KK states are always low lying although the
lowest part
of the spectrum is dominated by the Cartan if $R>>1$ or $R<<1$.
Thus, as we argued in the introduction, in the periods of time where
the radius is either large or small
the low energy spectrum is dominated by the Cartan, whereas in
between there is
a stringy region.
There is no topology change here.

In order to give a physical interpretation for these solutions we
have
to use the $\s$-model picture very carefully.
Consider a period in time where $R(t)\to 0,\infty$.
As we have seen in section 2, one of the radii in the $\s$-model
goes to zero and the spatial volume also goes to zero (qf.
(\ref{volume})).
One would say that during the period the ``universe" heads for a
``big crunch".
However, from the low energy point of view things seem different.
the inhabitants of this world see a large circular direction which
expands
(and becomes non-compact in the limit) but the excitations (momenta)
of the rest of the spatial dimensions are large so that they
effectively disappear
at low energy.
Thus it is a "big squeeze" but for reasons completely different from
those coming naively from the $\s$-model (see also \cite{os}).
That is, directions we thought were tiny, are not, and those we
thought
are finite, are not visible.
The effective string coupling in such periods becomes exponentially
small.

Once now we are in the regions where $R(t)\sim {\cal O}(1)$ there is
a unique
low energy scale set by $\alpha'$ (or equivalently $k$) and the low
energy
effective spectrum is that of a (deformed) $S^3$.
The string coupling here is finite, space dependent and can be made
almost everywhere arbitrarily small by adjusting $\phi_{0}$.

Thus we see that although the volume and $\s$-model geometry of the
``universe"
implies that we have a expanding or contracting universe, the low
energy
effective theory has a very different picture of it.
These  issues are very  important when we try to construct stringy
realistic cosmological models.

{\bf B}) Instead, we could have the same dependence as above for
$R(t)$ but
deforming $H^{3}_{+}$ now.
Here we have an example of dynamical topology change.
The change happens in the region around $R\sim 1$ as discussed in
section 2.
Similar remarks apply here as for the case discussed above.

{\bf C}) Case (ii) time dependence associated with deformed $SU(2)$
($\S(x)=\cos 2x$).

In this model the radius goes from $R=0$ at $t=0$ to $R=\infty$ at
$t=\pi\sqrt{k'/6}$.\footnote{It is very similar to the model
presented in \cite{nw}, and is in fact in the same $O(2,2,R)$-moduli
space of the $\s$-model \cite{gpp}.}
Similar remarks like in case A apply here. The only difference is
that here
we have an oscillating ``universe".
At  $t<<\sqrt{k'}$, $R(t)\sim t$, while around $t=\pi\sqrt{k'/6}$,
$R(t)\sim 1/(t-\pi\sqrt{k'/6})$.
Like (A), at the boundaries $R=0,\infty$ the effective string
coupling is finite and given by the dilaton of the SU(2)/U(1) model.
In between the model has an effective $S^3$ topology, but unlike
model (A)
the effective string coupling does not vanish at $R=1$ but is finite
and constant there.

Although this model appears (from the $\s$-model geometry point of
view) to
correspond to an eternally expanding and recontracting universe, the
real story is different, as in (A). The model starts out as a large
$S^1$ while the rest of space is curled up and turns into a
continuously deforming squashed
$S^3$ until it comes back to its original squeezed state.

{\bf D}) Case (ia,ib) time dependence
associated with deformed $SU(2)$ ($\S(x)=\cos 2x$). This case is
qualitatively similar to
(B) in the sense that the radius evolves from $0$ to $\infty$.
There is no periodicity in time however.

Another interpretation of the models above can be given.
If one considers the $SU(2)_{k}$ as a part of the compactified
dimensions of string theory, the model then provides a time-variation
of the masses (and other data like couplings) of the particle
spectrum in the adiabatic region.
This would describe a higher dimensional universe at $t=0$, where
some of its dimensions compactify as time go by, until their scale
becomes of the order of the Planck scale (this would correspond to
either the $tanh$ or the $coth$ solution (case (iii)).
In the context of the heterotic string, this would lead to late time
$SU(2)$ gauge symmetry.
This is a very interesting cosmological solution which in our opinion
deserves further study.

The stringy cosmological models presented above seem to have some
features which are not compatible with standard cosmological
scenarios (due to anisotropy etc).
It may be though, that isotropy is something that is achieved later
in the history of the universe\footnote{Some exact CFT models of that
sort
seem to be possible and will presented elsewere.}. In particular,
consider the product
of flat 2-d space times the $SL(2,R)_{k}/O(1,1)$ coset. In the first
of ref. \cite{nw} it was shown that for large times the spacetime was
isotropic.
The dual version of this model would give a universe where the Cartan
radius R of $SL(2,R)$ changes with t  as $R\sim t$. This would mean
that there are particles in this theory which see asymptotically an
isotropic space.

The solutions we presented  have some extra interesting features,
apart from the fact that being exact string solutions
they can give us a chance to understand stringy phenomena in the
context of cosmology.
In particular, although anisotropic inflation has not been analysed
so far we can remark that what plays here the role of the scale
factor is
$R(t)$ and there are regimes where it satisfies the usual conditions
for inflation, $R',R''>0$, namely, cases (ib,ii) and the $tanh$
branch of (iii).
It remains however to be seen if such conditions remain valid in the
presence of anisotropy.
Many interesting isotropic $\s$-model solutions exist, however we
don't know at present if they can be made exact solutions. This is a
serious drawback since
some of the interesting physics is supposed to happen at strong
curvatures
where our confidence on the solution breaks down, \cite{bv}.

As was mentioned in the introduction, another interesting application
of the Euclidean version of (\ref{s1},\ref{s2}), is the
identification of the angle $\th$ with Euclidean time. In this case
$R(t)$\footnote{Remember that now $t$ is a spatial coordinate.} plays
the role of the inverse temperature (Matsubara interpretation) which
varies spatially.
In this context, we have a background describing a thermal ensemble
of strings with a spatially non-uniform temperature. More precisely
there is
a temperature gradient in the $t$ direction.
Studying further this kind of solutions might give us useful
information
about Hagedorn-like instabilities and phase transitions in some
regions of
space.

Finally, some generalizations of the solutions above can be envisaged
using the same ideas as those discussed in this paper.
One would be higher dimensional generalizations of the above
providing
interesting cosmological solutions where some non-compact dimensions
compactify.
The other, would be to find all paths inside $O(d,d)$ which provide
exact string solutions. Looking at the boundaries  of $O(d,d)$ moduli
space may be the hint for finding such solutions.

\section{Exact wavelike solutions in string theory}

In this section we will discuss another class of exact solutions to
string theory which are time dependent and contrary to the solutions
presented above possess only a single scale, $\alpha'$.
They are closer in this respect to flat space.
We will focus in a particular solution of this kind \cite{w11}
which has high symmetry (it is a WZW model for a non-compact
non-semisimple group).
This solution can be interpreted as a
plane
gravitational (as well as $B_{\mu\nu}$) wave. It has seven Killing
symmetries and the corresponding 2-d $\sigma$-model
is
the WZW model for the group, $E^{c}_{2}$, the central
extension of the 2-d Euclidean group.
This model is interesting because it provides us with an exact
classical solution to string theory which has at the same time:

$\bullet$ a simple and clear physical interpretation,

$\bullet$ a non-trivial spacetime,

$\bullet$ it can be solved,

$\bullet$ it has unusual features, in particular the spacetime is
nowhere
asymptotically flat. Thus, a priori, it is not obvious that a
sensible
scattering matrix exists in such a background.

In \cite{kk1,kk2} we began solving the model by developing the
representation theory of the $E^{c}_{2}$ current algebra. In
particular, among other
things, we mapped the current algebra and the representation theory
to (almost) free fields.

We are interested in considering this background as an
exact classical solution to superstring theory.
Thus we will supersymmetrize the current algebra and we will again
map
the model to free bosons and fermions.
By studying the $\s$-model we will show
that the model has a $N=4$
worldsheet supersymmetry. The associated string
theory has a large unbroken spacetime supersymmetry.
For the type II string this is  N=4 spacetime supersymmetry.
In fact, this type of 4-d background turns out to
be a special case of the class of solutions found recently in
\cite{kkl}.
Moreover, we will explicitly construct the N=4 superconformal algebra
out of the (super) current algebra.

Some of the features though of the model like the potential spectrum
and and tree scattering of bosonic states have little difference
between the bosonic and the supersymmetric model. Thus we will find
the exact spectrum of the associated bosonic string theory
and we will compute the (modular invariant) vaccum energy.
There are two types of states in the theory, corresponding to
different kinds
of representations of the current algebra. We will qualitatively
describe their scattering.

We will start our discussion by presenting the current algebra
symmetry
of the background \cite{w11}.
This is specified by the OPEs
\be
J_{a}(z)J_{b}(w)={G_{ab}\over (z-w)^2}+{f_{ab}}^{c}{J_{c}(w)\over
(z-w)}+
\rr,\label{1}
\ee
where $(J_{1},J_{2},J_{3},J_{4})\sim(P_{1},P_{2},J,T)$, the $P_{i}$
generating the translations and $J$ being the generator of rotations
and $T$ being the central element.
The only non-zero structure constants are ${f_{31}}^{2}=1$
$G_{ab}$ is an invariant bilinear form (metric) of the algebra. $G$
is a symmetric matrix
and the Jacobi identities
constrain it so that $f_{abc}\equiv {f_{ab}}^{d}G_{dc}$ is completely
antisymmetric.
The most general invariant bilinear form for $E^{c}_{2}$ is
\be
G=k\left(\matrix{1&0&0&0\cr 0&1&0&0\cr 0&0&b&1\cr
0&0&1&0\cr}\right).
\ee
where we can assume without loss of generality that $k>0,b>0$
\cite{kk1}.
For $\hat E^{c}_{2}$ there is a unique solution to the Master
Equation \cite{HK}, which has
all the
properties of the Affine Sugawara construction. It is given by:
\be
L_{AS}={1\over 2k}\left(\matrix{1&0&0&0\cr 0&1&0&0\cr 0&0&0&1\cr
0&0&1&-b+{1\over k}\cr}\right)={1\over 2}G^{-1}+{1\over 2k^2}
\left(\matrix{0&0&0&0\cr 0&0&0&0\cr 0&0&0&0\cr
0&0&0&1\cr}\right),\label{3}
\ee

The $\s$-model realizing the current algebra above can be constructed
in a straightforward fashion, and provides the tool to study string
propagation.
Its action is \cite{w11}
\be
S={k\over 2\pi}\int d^{2}z\left[\d a_{i}\db a_{i}+\d u\db v+\db u\d
v+b\d u\db u-\e^{ij}a_{i}\db a_{j}\d u\right]\label{4}
\ee

{}From now on we will set $b=0$ by an appropriate shift in $v$. As
was
shown in \cite{kk1}, from the exact current algebra point of view,
once $b$ is finite
it can always rescaled away (assuming that the light-cone coordinates
are
non-compact).
We can also scale $k\to 1$ by appropriate scaling of
$a_{i},v$\footnote{
Effectively we are setting the parameter $Q$ of \cite{kk1} to one.}.
In this coordinate system, the nature of the background is not
obvious, however
as we will show below, it is in this background that the free field
representation of the algebra described in \cite{kk1} is manifest.
It should be also noted that in (4) the last term describes the
departure of the background from flat Minkowski space, and its
coupling can be made arbitrarily
small by a boost of the $u,v$ coordinates.
However, this does not imply that the perturbation is insignificant
since
the perturbing operator has bad IR behaviour.
It should also be mentioned here that the action (4) can be obtained
by an
$O(3,3,R)$ transformation of flat space.

The coordinate system where the nature of the background is more
transparent is given by \cite{w11}
\be
a_{1}=x_{1}+\cos u x_{2}\;\;,\;\; a_{2}=\sin u x_{2}\;\;,\;\;
v\to v+{1\over 2}x_{1}x_{2}\sin u\label{5}
\ee
where the action (\ref{4}) becomes
\be
S={1\over 2\pi}\int d^2 z\left[\d x_{i}\db x_{i}+2\cos u\d x_{2}\db
x_{1}+\d u\db v+\db u\d v\right].
\ee
In this form we can immediately identify the perturbation from flat
space
with a graviton+antisymmetric tensor mode given by $\cos u\d x_{2}\db
x_{1}$.
This is an operator with no IR problems, however its coupling is of
order one and cannot be rescaled at will.
One can however look at the structure of perturbation theory around
the flat background, and verify that indeed the current algebra
structure decribed in \cite{kk1} indeed emerges.

Before continuing further into the quantum theory, we shound point
out that
the string geodesics (solutions to the classical equations of motion
of the $\s$-model can be automatically obtained, since the model is
just a WZW model.
The solutions are obtained by writing the group elemnt as a matrix
product of a left moving and a right moving group element.

In \cite{kk1} we described a resolution of the $\ec$ current algebra
in terms
of free fields and studied its irreducible representations with
unitary base.
We will focus for the moment on one copy of the current algebra, and
return to
the full theory in due time.
We introduce four free fields, $x^{\mu}$ with
\be
\langle x^{\mu}(z)x^{\nu}(w)\rangle=\eta^{\mu\nu}\log(z-w)\;\;,\;\;
\eta^{\mu\nu}\sim\left(\matrix{1&0&0&0\cr
0&-1&0&0\cr 0&0&-1&0\cr 0&0&0&-1\cr}\right)
\ee
and define the light-cone $x^{\pm}=x^{0}\pm x^{3}$ and transverse
space
$x=x^{1}+ix^{2},\bar x=x^{1}-ix^{2}$ coordinates.
Then, the current algebra generators can be uniquely (up to Lorentz
transformations) written as
\be
J={1\over 2}\d x^{+}\;\;\;,\;\;\;T=\d x^{-}
\ee
\be
P^{+}=P^{1}+iP^{2}=ie^{-ix^{-}}\d
x\;\;\;,\;\;\;P^{-}=P^{1}-iP^{2}=ie^{ix^{-}}\d \xb
\ee
while the affine-Sugawara stress tensor (\ref{3}) becomes the free
field
stress
tensor
\be
T_{AS}={1\over 2}\eta_{\mu\nu}\d x^{\mu}\d x^{\nu}.
\ee

The current algebra representations can be built by starting with
unitary
representations of the $\ec$ algebra as a base and then
acting on them by the negative modes of the currents.
They fall into two classes:

(I) Representations with neither highest nor lowest weight
state.\footnote{
Except when $\vec p=0$ in which case the representations are one
dimensional.
}
The base is generated by vertex operators with $p_{-}=0$.
The representation is characterized by $p_{+}\in [0,1)$ and
transverse momentum $\px_{T}$.
The base operators are
\be
V^{I}_{p_{+},n}=e^{i(p_{+}+n)x^{-}+i\px_{T} \cdot \vec
x}.
\ee

(II) Representations with a highest weight state in the base.
Their conjugates have a lowest weight.
They have $p_{-}>0$ (negative for the conjugates).
The highest weight operator can be written as
\be
V^{II}\sim e^{ip_{+}x^{-}+ip_{-}x^{+}}H_{p_{-}}
\ee
where $H_{p_{-}}$ is the generating twist field in the transverse
space
corresponding to the transformation
\be
x(e^{2\pi i}z)=e^{-4\pi ip_{-}}x(z)\;\;,\;\;\bar
x(e^{2\pi i}z)=e^{4\pi ip_{-}}\xb(z).
\ee
The conformal weight of the operators at the base is
$\Delta=-2p_{+}p_{-}+p_{-}(1-2p_{-})$.

As we will see later on all the representations participate in the
modular invariant vacuum energy.
Thus we can easily describe the spectrum of physical states.
We will assume that the extra 22 dimensions are non-compact although
the compact case is as easy to handle.

$\bullet$ For type I states ($p_{-}=0$) we have the physical state
conditions
$L_{0}=\bar L_{0}=1$ which translate  to
\be
N=\bar N\;\;\;{\rm and}\;\;\;\vec
p_{T}^2=2-2N\;\;\;N=0,1,2,\cdots
\ee
where, $N$,$\bar N$ are the contribution of the transverse currents
or oscillators (from the left or right) and $\vec p_{T}$ is the 24-d
transverse momentum.
It is obvious that for $N=0$ we have $\vec p_{T}^2=2$ which is a
component
of the tachyon, while for $N=1$ we have $\vec p=0$ and we obtain some
modes
of the massless fields (graviton, antisymmetric tensor and dilaton).
which propagate along the wave.
For $N>1$ there are no physical states.

$\bullet$. For type II states the situation is like in usual string
theory,
it is just their dispersion relation that changes:
\be
-4p_{+}p_{-}+2p_{-}(1-2p_{-})+\vec p_{T}^2=2-2N\label{15}
\ee
and $\vec p_{T}$ now stands for the 22 transverse dimensions.
Eq. (\ref{15}) can also be written after $p_{-}\to p_{-}/4$,
$p_{+}\to
p_{+}-p_{-}/4+1/2$ as
\be
p_{+}p_{-}-\vec p_{T}^{2}=2(N-1)
\ee
which is the flat space spectrum but in 24-d Minkowski space.
Thus for almost all states the wave reduces the effective
dimensionality by two. As we will see below, this happens because the
states are localized in the extra two dimensions.

Eventually we would lie to compute the partition function (vaccuum
energy).
In order to do this we can guess the modular invariant sowing of
representations by looking at the wavefunctions on the group.
These wavefunctions can be computed
and we easily deduce that we have the standard diagonal modular
invariant.

{}From these wavefunctions we can see the qualitative behaviour of
the fluctuations in this background.
Type I states are just plane waves. As for type II states,
if we go to the $x_{i}$ coordinates (\ref{5}) where the wave nature
is
manifest,
then we see that the states are localized where $x^{2}_{1}+x^{2}_{2}+
2x_{1}x_{2}\cos(u)=0$. So for fixed time the state is localized  on
the two
lines $x_1=\pm x_{2}$, $x_3=$constant where $u=t-x_{3}$.
This disturbance travels to the left in the $x_3$ direction with the
speed of light (in flat space).

We can now try to construct the vaccuum energy by putting together
the representations.
What can happen at most is the truncation phenomenon as in the
compact case.
In trying to built the modular invariant partition function we have
to overcome
the problem that the representations are infinite dimensional and
since all the states in the base have the same energy the partition
function diverges.
We will have to regulate this divergence and ensure that upon the
removal
of the regulator we will obtain a modular invariant answer.
We will start by computing the character formulae for the
representations above.
In fact what we are interested in is the so called signature
character which keeps track of the positive and negative norms.
We will not worry much about the type I representations (although
they are the easiest to deal with) since their contribution is of
measure zero.
For the type II representations we can easily compute the signature
character
\be
\chi^{II}_{p_{+},p_{-}}(q,w)\equiv
Tr[(-1)^{\epsilon}q^{L_{0}-{c\over 24}}w^{J_{0}}]
\ee
where $\epsilon =0,1$ for positive (respectively negative) norm
states.
If $2p_{-}\not=$integer\footnote{We can also compute the character
when $2p_{-}\in Z$ but this is not needed again for the partition
functions since it is of measure zero.} then the only non-trivial
null vector is the one at the
base responsible for the fact that we have a highest (or lowest)
weight representation. Thus the character is \cite{BK}
\be
\chi^{II}_{p_{+},p_{-}}(q,w)={q^{-2p_{+}p_{-}+p_{-}(1-2p_{-})
-{1\over 6}}
w^{p_{+}}\over
(1-w)\prod_{n=1}^{\infty}(1-q^{n}w)(1-q^{n}w^{-1})}=\nonumber
\ee
\be
=iq^{-2p_{+}p_{-}+p_{-}(1-2p_{-})-{1\over 12}}w^{p_{+}-{1\over 2}}
{\eta(q)\over \vartheta_{1}(v,q)}
\ee
where $\vartheta_{1}$ is the standard Jacobi $\vartheta$-function and
$w=e^{2\pi i v}$.
The infinity we mentioned before can be seen as $w\to 1$ as the pole
in the $\vartheta$-function.
In order to see how we must treat this infinity we will calculate
this partition function in the quantum mechanical case (that is
keeping
track only of the zero modes).
We will introduce a twisted version of the (Minkowski) amplitute for
the particle to go from $x$ to $x'$
in time $\tau$:
\be
<x|x';\tau,v,\bar v>=<x|e^{i\tau H}e^{\zeta
J_{0}+\bar \zeta\bar
J_{0}}|x'>\label{19}
\ee
The quantum mechanical partition function can be obtained by setting
$x=x'$, $\zeta,\bar \zeta\to 0$ and integrating over $x$.
This last integral gives the overall volume of space that we will
drop
since we are interested in the vaccuum energy per unit volume.
The amplitude (\ref{19}) in the coordinates $u,v$ and
$a_{1}=r\cos\theta,a_{2}=r\sin\theta$, can be
calculated with the result
\be
<x|x';\tau,\zeta,\bar \zeta>\sim {1\over \tau^2}{(u-u'+\bar
\zeta-\zeta)\over \sin\left(
{u-u'+\bar \zeta-\zeta\over 2}\right)}\exp\left[i{(u-u'+\bar
\zeta-\zeta)^2\over 4\tau}-i
{(u-u'+\bar \zeta-\zeta)(v-v')\over 2\tau}-\right.\nonumber
\ee
\be
-\left.{i\over 8\tau}(u-u'+\bar \zeta-\zeta)
\cot\left({u-u'+\bar \zeta-\zeta\over
2}\right)\left(r^2+r'^2-2rr'{\cos\left(
\theta-\theta'-{u-u'+\zeta+\bar \zeta\over 2}\right)\over
\cos\left({u-u'+\bar \zeta-\zeta\over 2}\right)}\right)\right]
\ee
We can then perform the limits to obtain the $finite$ result
\be
Z(\tau)=<x|x;\tau,\zeta=0,\bar \zeta=0>\sim {1\over
\tau^{2}}\label{20}
\ee
This result may seem surprising since we have only two continuous
components of the momentum, namely $p_{+}$ and $p_{-}$ but no
transverse
momentum.
However the result is not as surprising as it looks at first, and to
persuade the reader  to that we will provide with an analogous
situation where the answer is obvious.
Consider the case of a flat 2-d plane. The zero mode spectrum are the
usual plane waves, $e^{i\vec p\cdot \vec x}$, but we will work in the
rotational
basis, where the appropriate eigenfunctions corresponding to energy
$p^2$
are $e^{im\theta}J_{|m|}(pr)$.
In such a basis, we will have precisely the same problem in
calculating
the partition function as we had above. Namely, for a given $p$ there
are an infinite number of states (numbered by $m\in Z$) with the same
energy.
Thus the partition function seems to be infinite.
However here we know what to do: go to the (good) plane wave basis,
or
calculate the propagator and then take the points to coincide in
order
to get the partition function. This will also give the right result.
This is precisely the prescription we have applied above and we found
that
the quantum mechanical partition function of the gravitational wave
zero modes is precisely that of flat Minkowski space.
Having found the way to sum the zero mode spectrum, it is not
difficult to show
using (\ref{20}) that the vaccum energy of the associated bosonic
string
theory
is equal to the flat case, namely
\be
F=\int{d\tau d\bar \tau\over Im\tau^2}(\sqrt{Im\tau}\eta\bar
\eta)^{-24}
\ee
where we have added another 22 flat non-compact dimensions\footnote{
The extra 22 dimensions could be compactified with no additional
effort}.

We would add here a few comments about scattering. We hope to present
the full
picture in the future.
States corresponding to type I representations scatter among
themselves, this is already obvious from their vertex operator
expressions \cite{kk1}.
This is dangerous however, since we might have trouble with
unitarity.
We will check here that in a 4-point amplitude of type I tachyons
only
physical states appear as intermediate states.
Remember that type I tachyons have $p_{-}=0$. Thus the four type I
tachyons are characterized  by their $p_{+}^{i}$ and $\vec
p_{T}^{i}$.
The 4-point amplitude is then given by the standard
$\delta$-functions
of $p_{+}$ and $\vec p_{T}$ multiplied by the Shapiro-Virasoro
amplitude
with one difference: instead of the invariants $p_{i}\cdot p_{j}$ we
now have them restricted to transverse space, $\vec p^{i}_{T}\cdot
\vec
p^{j}_{T}$.
\setcounter{footnote}{0}
This is similar to the Shapiro-Virasoro amplitude in 24 Euclidean
dimensions
(modulo the delta functions)
and its analytic structure is different.
It can be easily checked that instead of the infinite sequence of
poles
of the usual amplitude here we have just two poles, one corresponding
to intermediate on-shell tachyons and the other to intermediate
on-shell massless type I particles. This is in agreement with the
fusion rules of the current algebra.\footnote{Scattering for type I
states
for arbitrary plane waves has been considered in \cite{r1} with
similar results.}

The scattering of the type II states is more complicated and will be
dealt with elsewhere.

The bosonic string vacuum  as it stands is ill defined in the quantum
theory due to the presence of the tachyon in its spectum.
We will construct now the N=1 extension of this background, in order
to
make a superstring out of it.

In order to do this we  will add four free fermions (with Minkowski
signature),
$\psi_{1,2}$, $\psi_{J}$, $\psi_{T}$, normalized as
\be
\psi_{a}(z)\psi_{b}(w)={G_{ab}\over (z-w)}+\rr
\ee
The N=1 supercurrent is given by
\be
G=E^{ab}J_{a}\p_{b}-{1\over 6}f^{abc}\p_{a}\p_{b}\p_{c}
\ee
where
\be
E={1\over k}\left(\matrix{1&0&0&0\cr 0&1&0&0\cr 0&0&0&1\cr
0&0&1&-b+1/2k\cr}\right)\label{24}
\ee
and the group indices are always raised and lowered with the
invariant metric $G_{ab}$.
The only non-zero component of $f^{abc}$ is $f^{124}=1/k^2$.
It can be verified that (\ref{24}) satisfies the superconformal ME
\cite{GHKO}
and preserves the current algebra structure.
In particular
\be
G(z)G(w)={2c/3\over (z-w)^3}+{2T(w)\over (z-w)}+\rr
\ee
with $c=6$, $T=T^{b}+T^{f}$, with $T^{b}$ being the affine Sugawara
stress tensor and $T^{f}$ the free fermionic stress tensor
\be
T^{f}=-{1\over 2}G^{ab}\p_{a}\d \p_{b}
\ee
$G$ is also primary with conformal weight $3/2$ with respect to $T$.
The supersymmetric currents in the theory are the superpartners of
the free
fermions
\be
G(z)\p_{a}(w)={\hat J_{a}(w)\over (z-w)}+\rr
\ee
It is easy to find that
\be
\hat P_{1}=P_{1}-{1\over k}\p_{2}\p_{T}\;\;,\;\;\hat
P_{2}=P_{2}+{1\over k}\p_{1}\p_{T}
\ee
\be
\hat J=J+{1\over 2k}T-{1\over k}\p_{1}\p_{2}\;\;,\;\;\hat
T=T
\ee
The supersymmetric currents $\hat J_{a}$ satisfy the same algebra
(\ref{1}) as the bosonic ones $J_{a}$. This should be contrasted with
the
compact case
where the level is shifted by the dual Coxeter number.
The supercurrent $G$ has a similar expression in terms of $\hat
J_{a}$ as in the compact case
\be
G=G^{ab}\hat J_{a}\p_{b}-{1\over
3}f^{abc}\p_{a}\p_{b}\p_{c}\ee

As in the bosonic case, the supersymmetric current algebra
can be written in terms of free bosons, $x^{\pm},x,\bar x$ and
fermions
$\p^{\pm},\p,\pp$ with
\be
<x^{+}(z)x^{-}(w)>=-<x(z)\bar x(w)>=2\log(z-w)
\ee
\be
<\p(z)\pp(w)>=-<\p^{+}(z)\p^{-}(w)>={2\over (z-w)}
\ee
all others being zero.
Explicitly,
\be
P_{1}+iP_{2}=i\sqrt{k}e^{-ix^{-}/Q}\d
x\;\;,\;\;P_{1}-iP_{2}=i\sqrt{k}e^{ix^{-}/Q}\d \bar x
\ee
\be
J={Q\over 2}\d x^{+}+{1\over 2}\left(Q+{1\over Q}\right)\d x^{-}
+{i\over 2}\p\pp\;\;,\;\;T={Q\over b}\d x^{-}
\ee
\be
\p_{1}+i\p_{2}=\sqrt{k}e^{-ix^{-}/Q}\p\;\;,\;\;\p_{1}-i\p_{2}=
\sqrt{k}e^{ix^{-}/Q}\pp
\ee
\be
\p_{J}=i{Q\over 2}\left(\p^{+}+\p^{-}\right)\;\;,\;\;\p_{T}=i
{Q\over b}\p^{-}
\ee
where $Q=\sqrt{kb}$.
The fermionic admixture to $J$ is added to guarantee that $J_{a}$ and
$\p_{a}$
commute.

It is not difficult to see that the stress tensor and supercurrent
are
free in the free-field basis
\be
T={1\over 2}\left[\d x^{+}\d x^{-}-\d x \d \bar x\right]+{1\over
4}\left[
\p^{+}\d \p^{-}+\p^{-}\d \p^{+}-\p\d \pp-\pp\d\p\right]\label{37}
\ee
\be
G={i\over 2}\left[ \d x\pp+\d \bar x\p+\d x^{-}\p^{+}+\d
x^{+}\p^{-}\right]
\ee
In the free field basis it can be easily seen that the theory has an
N=2 superconformal algebra. The conventionally normalized U(1)
current that determines the "complex
structure" is given by
\be
\j={1\over 2}\left[\p\pp-\p^{+}\p^{-}\right]
\ee
whereas the second supercurrent is
\be
G^{2}=-{1\over 2}\left[\d x \pp-\d \bar x\p+\d x^{+}\p^{-}-\d
x^{-}\p^{+}\right]\ee

In conformity with our $\s$-model discussion later , we will also
consider
the presence of the dilaton, which here is reflected as background
charge
in the lightcone directions.
We have the following modifications for the N=2 generators
\be
\delta\j=-i\left(Q^{+}\d x^{-}+Q^{-}\d
x^{+}\right]\;\;\;,\;\;\;\delta T={i\over
2}\left(Q^{+}\d^{2}x^{-}-Q^{-}\d^{2}x^{+}\right)
\ee
\be
\delta G^{1}=Q^{+}\d \p^{-}-Q^{-}\d \p^{+}\;\;\;,\;\;\;\delta
G^{2}=i\left(Q^{+}\d \p^{-}+Q^{-}\d \p^{+}\right)
\ee
and $\delta c=-8Q^{+}Q^{-}$.

The final step is the realization that when $Q^{-}=0$ the theory has
even larger superconformal symmetry, namely N=4.
The N=4 superconformal algebra in question contains the stress
tensor, $SU(2)_{k}$ currents $J^{a}$  and four supercurrents that
transform as conjugate doublets under the $SU(2)$.
It is defined in terms of the OPEs
\be
J^{a}(z)J^{b}(w)={k\over 2}{\delta^{ab}\over
(z-w)^2}+i\epsilon^{abc}{J^{c}(w)\over (z-w)}+\rr
\ee
\be
J^{a}(z)G^{i}(w)={1\over 2}\s^{a}_{ji}{G^{j}(w)\over
(z-w)}+\rr\;\;,\;\;
J^{a}(z)\bar G^{i}(w)=-{1\over 2}\s^{a}_{ij}{\bar G^{j}(w)\over
(z-w)}+\rr\ee

\be
G^{i}(z)\bar G^{j}(w)={4k\delta^{ij}\over
(z-w)^3}+2\s^{a}_{ji}\left[
{2J^{a}(w)\over (z-w)^2}+{\d J^{a}(w)\over
(z-w)}\right]+2\delta^{ij}{
T(w)\over (z-w)}+\rr\ee
the rest being regular.
$J^{a}$, $G^{i},\bar G^{i}$ are primary with the appropriate
conformal weight
and $c=6k$.
In our case the SU(2) current algebra has level $k=1$.
The N=4 generators, in the free field basis are
\be
G^{1}={i\over \sqrt{2}}\left[\d\bar x\p+\d
x^{-}\p^{+}\right]\;\;,\;\;
G^{2}=-{i\over \sqrt{2}}\left[\d \bar x\p^{-}+\d x^{-}\pp\right]
e^{iQ^{+}x^{-}}
\ee

\be
\bar G^{1}={i\over \sqrt{2}}\left[\d x \pp+\d
x^{+}\p^{-}-2iQ^{+}\d\p^{-}\right],\bar G^{2}={i\over \sqrt{2}}\left[
\d x \p^{+}+\d x^{+}\p
-iQ^{+}\p\p^{+}\p^{-}\right]e^{-iQ^{+}x^{-}}\ee
\be
J^{1}+iJ^{2}={1\over 2}\p\p^{+}e^{-iQ^{+}x^{-}}\;\;,\;\;
J^{1}-iJ^{2}={1\over 2}\pp\p^{-}e^{iQ^{+}x^{-}}\ee
\be
J^{3}={1\over 4}\left[\p\pp-\p^{+}\p^{-}-2iQ^{+}\d
x^{-}\right]\ee
and $T$ is given in (\ref{37}).

One can invert the map to free fields and right these operators in
terms of the original ($\s$-model) currents. Care should be exercized
though in normal ordered expressions.
What we need here is the appearence, in the N=4 generators, of terms
which are non-local in the original currents.
These terms are of the form $exp\left[\pm{i\over k}(QQ^{+}-1)\int T
dz\right]$.
However, in the sigma model, the current $T$ is given as a total
derivative of
a free field, $T=\d u$, with $\d\db u=0$.
Thus the exponential factors are just $exp\left[\pm{i\over
k}(QQ^{+}-1)u(z)\right]$.
Notice also that they dissapear at a specific value of the background
charge, $Q^{+}=1/Q$.

To conclude this section, exact solutions as the one above, have very
interesting dynamics, and they will sharpen our understanding of
strings propagating in non-trivial background fields.
Several issues remain open however, the most interesting in our
opinion being to compute the scattering of the bulk of he spectrum in
the supersymmetric case.

\noindent
{\bf Acknowledgments}

\noindent
The work of C. Kounnas was
partially supported by EEC grants SC1$^{*}$-0394C and
SC1$^{*}$-CT92-0789.

\end{document}